\documentclass[aps,prd,twocolumn,showpacs,nofootinbib]{revtex4}

\usepackage{graphicx}% Include figure files
\usepackage{dcolumn}% Align table columns on decimal point

\begin{document}

% Use the \preprint command to place your local institutional report
% number in the upper righthand corner of the title page in preprint mode.
% Multiple \preprint commands are allowed.
% Use the 'preprintnumbers' class option to override journal defaults
% to display numbers if necessary
%\preprint{}

%Title of paper
\title{Ladder Dyson--Schwinger calculation of the anomalous
{\boldmath $\gamma$-$3\pi $\unboldmath} form factor}

% repeat the \author .. \affiliation  etc. as needed
% \email, \thanks, \homepage, \altaffiliation all apply to the current
% author. Explanatory text should go in the []'s, actual e-mail
% address or url should go in the {}'s for \email and \homepage.
% Please use the appropriate macro foreach each type of information

% \affiliation command applies to all authors since the last
% \affiliation command. The \affiliation command should follow the
% other information
% \affiliation can be followed by \email, \homepage, \thanks as well.
\author{Stephen R. Cotanch}
\email[]{cotanch@ncsu.edu}
%\homepage[]{Your web page}
%\thanks{}

\author{Pieter Maris}
\email[]{pmaris@unity.ncsu.edu}
%\homepage[]{Your web page}
%\thanks{}
\affiliation{Department of Physics,
North Carolina State University, Raleigh,  NC 27695-8202}

\date{\today}

\begin{abstract}
The anomalous processes, $\gamma \to 3 \pi$ and $\gamma\pi
\to \pi\pi$, are investigated within the Dyson--Schwinger
framework using the rainbow-ladder approximation.  Calculations reveal
that a complete set of ladder diagrams beyond the impulse
approximation are necessary to reproduce the fundamental low-energy
theorem for the anomalous form factor.  Higher momentum calculations
also agree with the limited form factor data and exhibit the same
resonance behavior as the phenomenological vector meson
dominance model.
\end{abstract}

% insert suggested PACS numbers in braces on next line

% insert suggested keywords - APS authors don't need to do this
%\keywords{}

%\maketitle must follow title, authors, abstract, \pacs, and \keywords
\maketitle

%%%%%%%%%%%%%%%%%%%%%%%%%%%%%%%%%%%%%%%%%%%%%%%%%%%%%%%%%%%%%%%%%%%%%%%%%%%%%
%
\section{\label{sec:intro}
Introduction}
Anomalies, in particular the anomalous nonconservation of the axial
vector current, are a consequence of renormalization and are thus
potentially present in all gauge theories.  In fact a conserved axial
current is incompatible with electromagnetic gauge invariance since
radiative corrections lead to a non-zero current divergence.  The
quintessential example is the $\pi^0 \to \gamma \gamma$ decay, which,
in the chiral limit, would be forbidden if the axial current was
conserved.  However, the form factor for this process,
$F^{2\gamma}(Q^2)$, is non-zero and in the combined chiral and
zero-momentum limit it is given by $F^{2\gamma}(Q^2) \to {\hat
F}^{2\gamma}(0) = e^2/(4\pi^2 {\hat f}_\pi) = \alpha/(\pi {\hat
f}_\pi)$~\cite{Adler:1969gk,Bell:1969ts}, where $e$ is the proton
charge, $\alpha$ is the fine structure constant and ${\hat f}_\pi$ is
the pion decay constant in the chiral limit.

This work addresses another anomalous process, $\gamma \to \pi^+ \pi^0
\pi^-$, which also should, but doesn't, vanish in the combined chiral
and soft momentum limits.  The form factor for this, and the crossing
related process $\gamma\pi \to \pi\pi$, is denoted by
$F^{3\pi}(s,t,u)$, where $s = -(P_3+P_4)^2$, $t = -(P_2+P_4)^2$, and
$u = -(P_2+P_3)^2$, with $P_i$ the pion momenta.  In the chiral limit,
and for zero four-momenta,
Refs.~\cite{Adler:1971nq,Terent'ev:1972kt,Aviv:1972hq} proved
${\hat F}^{3\pi}(0,0,0) = {\hat F}^{2\gamma}/(e {\hat f}_\pi^2) =
e /(4 \pi^2 {\hat f}_\pi^3)$.

Several issues, both experimental and theoretical, motivate this
investigation.  Experimentally, only limited low momentum information
about $F^{3\pi}$ is available from Primakov production using the $\pi
A \to \pi\pi A$ reaction~\cite{Antipov:1987tp} and, for timelike
photons, from $e^+ e^- \to 3 \pi$
measurements~\cite{Dolinsky:1991vq,Antonelli:1992jx,Barkov:1987ca}.
However, measurements to determine the form factor for the process
$\gamma\pi \to \pi\pi$ are now underway at JLab~\cite{Miskimen:1994}.
New results for the form factor $F^{3\pi}$ in the range $0.27\;{\rm
GeV}^2 < s < 0.72\;{\rm GeV}^2$ are expected to be released in the
near future~\cite{Miskimen:private}.

Theoretically, there are published
analyses~\cite{Alkofer:1996jx,Bistrovic:1999dy,Bistrovic:1999yy} based
on the set of Dyson--Schwinger equations [DSEs] in the rainbow-ladder
truncation.  These pioneering studies evaluated diagrams in the
generalized impulse approximation [GIA].  However, it is necessary to
go beyond the impulse approximation in order to correctly describe
$\pi$-$\pi$ scattering~\cite{Bicudo:2001jq,Cotanch:2002vj}:
consistency requires inclusion of all planar diagrams ignoring gluon
self-interactions (triple and quartic gluon vertices).  In the current
study we apply the same scheme to $\gamma$-$3\pi$ processes and
elucidate that the rainbow-ladder truncation in combination with the
GIA is insufficient for a realistic description.

The structure of this paper spans six sections.  In the next section
the DSE formalism for mesons is applied to evaluate $F^{3\pi}$.  Our
numerical results for the form factor for symmetric pion momenta are
presented in Sec.~\ref{sec:symresults} and the low-energy theorem is
reproduced.  In Sec.~\ref{sec:mesonexchange} we compare our results
with meson exchange models for more general pion momenta.  Finally, we
confront the limited data in Sec.~\ref{sec:expdata} and then
summarize our conclusions in last section.

%%%%%%%%%%%%%%%%%%%%%%%%%%%%%%%%%%%%%%%%%%%%%%%%%%%%%%%%%%%%%%%%%%%%%%%%%%%%%
\section{\label{sec:mesons}
Meson interactions in the Dyson--Schwinger approach}

\subsection{\label{sec:DSEs}
Dyson--Schwinger equations}
Here we briefly summarize the DSE formalism applied to mesons.  For
more complete details consult
Refs.~\cite{Roberts:1994dr,Roberts:2000aa,Alkofer:2000wg,Maris:2003vk}.
Working in Euclidean space, with metric
\mbox{$\{\gamma_\mu,\gamma_\nu\} = 2\delta_{\mu\nu}$},
\mbox{$\gamma_\mu^\dagger = \gamma_\mu$} and \mbox{$a\cdot b = a_i b_i
\equiv \sum_{i=1}^4 a_i b_i$}, the DSE for the renormalized quark
propagator having four-momentum $p$ is
\begin{eqnarray}
 S(p)^{-1} &=& i \, Z_2\, /\!\!\!p + Z_4\,m_q(\mu) +
\nonumber \\ && {}
        Z_1 \int_q^\Lambda\! g^2 D_{\mu\nu}(k)
	        \textstyle{\frac{\lambda^{\alpha}}{2}}
 		\gamma_\mu S(q)\Gamma^{\alpha}_\nu(q,p) \,.
\label{gendse}
\end{eqnarray}
Here ${\alpha} = 1...8$ is the color index, $D_{\mu\nu}(k)$ the
dressed-gluon propagator and $\Gamma^{\alpha}_\nu(q,p)$ the
dressed-quark-gluon vertex.  The symbol $\int_q^\Lambda\!$ represents
$\int_q^\Lambda\!\frac{d^4k}{(2\pi)^4}$ with \mbox{$k=p-q$}.
Equation~(\ref{gendse}) has the general solution
\mbox{$S(p)^{-1} = i/\!\!\! p A(p^2) +$} \mbox{$B(p^2)$} which is
renormalized at spacelike $\mu^2$ so that \mbox{$A(\mu^2)=1$} and
\mbox{$B(\mu^2)=m_q(\mu)$} with $m_q(\mu)$ being the current quark
mass.

Mesons are modeled as $q^a \bar{q}^b$ bound states governed by the
Bethe--Salpeter amplitude [BSA], $\Gamma_H$, which is a solution of
the homogeneous Bethe--Salpeter equation [BSE].  For quark flavors
$a$, $b$ and incoming, outgoing momenta $p_+ = p + P/2$, $p_- = p-
P/2$, the BSE is
\begin{eqnarray}
 \Gamma^{a\bar{b}}_H(p_+,p_-) &=&
        \int_q^\Lambda\! K(p,q;P)
\nonumber \\ && {}
        \times S^a(q_+) \, \Gamma^{a\bar{b}}_H(q_+,q_-) \, S^b(q_-)\, .
\label{homBSE}
\end{eqnarray}
The momenta $q_\pm$ are similarly defined.  The renormalized,
amputated $q\bar q$ scattering kernel, $K$, is irreducible with
respect to a pair of $q\bar q$ lines.  Equation~(\ref{homBSE}) only
admits solutions for discrete values of $P^2 = -m_H^2$, where $m_H$ is
the meson mass.  Imposing the canonical normalization condition for
$q\bar q$ bound states then uniquely determines $\Gamma_H$.  For
pseudoscalar bound states the most general decomposition for
$\Gamma_{PS} \equiv \Gamma^{a\bar{b}}_H$ is~\cite{LS:1969,Maris:1997tm}
\begin{eqnarray}
\label{eq:piBSA}
\lefteqn{\Gamma_{PS}(q_+,q_-) = \gamma_5 \big[ i E(q^2,q\cdot P)
        + \;/\!\!\!\! P \, F(q^2,q\cdot P) }
\nonumber \\ && {}
        + \,/\!\!\!q \, G(q^2,q\cdot P)
        + \sigma_{\mu\nu}\,P^\mu q^\nu \,H(q^2,q\cdot P) \big]\,,
\end{eqnarray}
where the amplitudes $E$, $F$, $G$ and $H$ are Lorentz scalar
functions of $q^2$ and $q\cdot P$.  The odd C parity of the neutral
pion requires that the amplitude $G$ be odd in $q\cdot P$, while the
others are even.

\subsection{\label{sec:rainbowladder}
Rainbow-ladder truncation}
The central approximation in this work is the rainbow-ladder
truncation for the set of DSEs.  For the quark DSE,
Eq.~(\ref{gendse}), the rainbow truncation is
\begin{equation}
\label{ourDSEansatz}
Z_1 \, g^2 D_{\mu \nu}(k) \Gamma^{\alpha}_\nu(q,p) \to
 	{\cal G}(k^2) D_{\mu\nu}^{\rm free}(k)\, \gamma_\nu
	\textstyle{\frac{\lambda^{\alpha}}{2}} \,,
\end{equation}
where $D_{\mu\nu}^{\rm free}(k=p-q)$ is the free gluon propagator in
the Landau gauge and ${\cal G}(k^2)$ is an effective $\bar q q$
interaction that reduces to the perturbative QCD running coupling in
the ultraviolet region.  For the BSE, Eq.~(\ref{homBSE}), the ladder
truncation is
\begin{equation}
\label{ourBSEansatz}
        K(p,q;P) \to
        -{\cal G}(k^2)\, D_{\mu\nu}^{\rm free}(k)
        \textstyle{\frac{\lambda^{\alpha}}{2}}\gamma_\mu
        \textstyle{\frac{\lambda^{\alpha}}{2}}\gamma_\nu \,.
\end{equation}
These two truncations, in tandem, yield vector and axial vector
vertices satisfying their respective Ward--Takahashi identities.  This
ensures chiral symmetry is respected with the pion being a massless
Goldstone boson in the chiral
limit~\cite{Maris:1997tm,Delbourgo:1979,Bender:1996,Maris:1998hd}. Further,
a conserved electromagnetic current is also obtained if the GIA is
used to calculate electromagnetic form
factors~\cite{Roberts:1996,Maris:2000sk}.

\subsection{
Amplitude for {$\gamma$-$3\pi$} processes}
\begin{figure}[h]
\includegraphics[width=7cm]{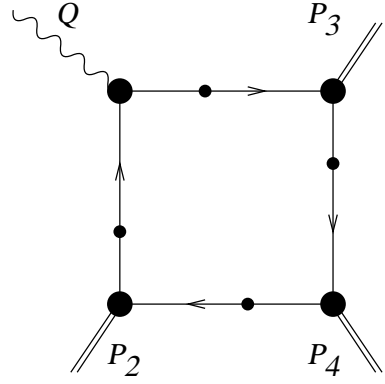}
\caption{\label{fig:GIA}
The generic GIA diagram for $\gamma$-$3\pi$ processes.
All external momenta flow inward.  The photon and three pions
have momenta $Q$ and $P_i$, $i=2,3,4$, respectively.}
\end{figure}

The above formalism is now applied to the $\gamma \to 3\pi$ process to
compute the anomalous amplitude, $A^{3\pi}$, and the corresponding
form factor, $F^{3\pi}$.  For simplicity, let us first consider one of
the configurations that contribute to this process, depicted in the
GIA in Fig.~\ref{fig:GIA}.
The quark lines in this figure represent dressed quark propagators,
obtained as solutions of their DSE.  The pion BSAs are solutions of
the homogeneous BSE, and also the quark-photon vertex is dressed.

The Lorentz structure for this diagram is
\begin{eqnarray}
 a_\mu(P_2, P_3, P_4) &=& -i\epsilon_{\mu\nu\rho\sigma}
		P_2^\nu P_3^\rho P_4^\sigma  \; f(s,t,u)\,,
\label{eq:Lora}
\end{eqnarray}
where $f(s,t,u)$ is a Lorentz scalar function of the Mandelstam
variables $s=-(Q+P_2)^2$, $t = -(Q+P_3)^2$, $u = -(Q+P_4)^2$.  All
three pions are on-shell, $P_i^2 = m_\pi^2$.  The photon momentum is
$Q= -(P_2 + P_3 + P_4)$ and $Q^2$ is related to the Mandelstam
variables via
\begin{eqnarray}
	s + t + u &=& 3 \, m_\pi^2 - Q^2  \,.
\label{eq:MandelQ2}
\end{eqnarray}

The total $\gamma \to 3\pi$ amplitude is obtained by summing over all
six permutations of the three pions with the appropriate charge
factors.  It has the same Lorentz structure as Eq.~(\ref{eq:Lora}),
and can be written as
\begin{eqnarray}
 A_\mu^{3\pi}(P_2, P_3, P_4) &=& -i\epsilon_{\mu\nu\rho\sigma}
		P_2^\nu P_3^\rho P_4^\sigma  \; F^{3\pi}(s,t,u) \,.
\end{eqnarray}
In the isospin limit (equal masses for up and down quarks and ignoring
electromagnetic corrections) all six configurations can be related
through symmetry operations, which yields
\begin{eqnarray}
\lefteqn{ F^{3\pi}(s,t,u) = }
\nonumber\\
 &&\frac{e}{3}\,
	\Big(f(s,t,u) + f(u,s,t) + f(t,u,s)\Big)\,.
\label{eq:perm}
\end{eqnarray}

The process $\gamma\pi \to \pi\pi$ is described by the same form
factor: the only difference is in the kinematical variables.  For
physical three-pion production, all three Mandelstam variables $s$,
$t$ and $u$ must be above the threshold value $4\,m_\pi^2$.  For
$\gamma\pi \to \pi\pi$, the kinematics are different: $s > 4\,m_\pi^2$
but $t,u < 0$.

\subsection{
Corrections to the GIA}
Although the GIA is consistent with current conservation for
three-body processes such as electromagnetic form factors, it is
insufficient, as demonstrated in detail in the next section, for
four-body amplitudes such as $\pi$-$\pi$ scattering and $F^{3\pi}$.
It was previously noted~\cite{Alkofer:1996jx} that if the Ball--Chiu [BC]
ansatz~\cite{Ball:1980ay} is used for the quark-photon vertex in
combination with
\begin{eqnarray}
 \Gamma_{PS} = \frac{i\gamma_5}{f_\pi} B(q^2) \ ,
\end{eqnarray}
the GIA does reproduce the low-energy theorem~\footnote{
Note that the pion decay constant depends on the current quark mass:
the chiral limit value is a few percent smaller than the physical
value, $f_\pi=92.4~{\rm MeV}$.  In the model used here
${\hat f}_\pi=90~{\rm MeV}$.}
\begin{eqnarray}
 {\hat F}^{3\pi}(0,0,0) &=&
	\frac{e}{4\,\pi^2\,{\hat f}_\pi^3}
		\approx 10.5 \, \text{GeV}^{-3}  \ ,
\label{eq:chiralanomaly}
\end{eqnarray}
independent of model details  for the scalar part of the
quark self-energy, $B(q^2)$.  However, this
result is lost when the complete pion BSA (see Eq.~(\ref{eq:piBSA}))
is used.

\begin{figure}[h]
\includegraphics[width=8.5cm]{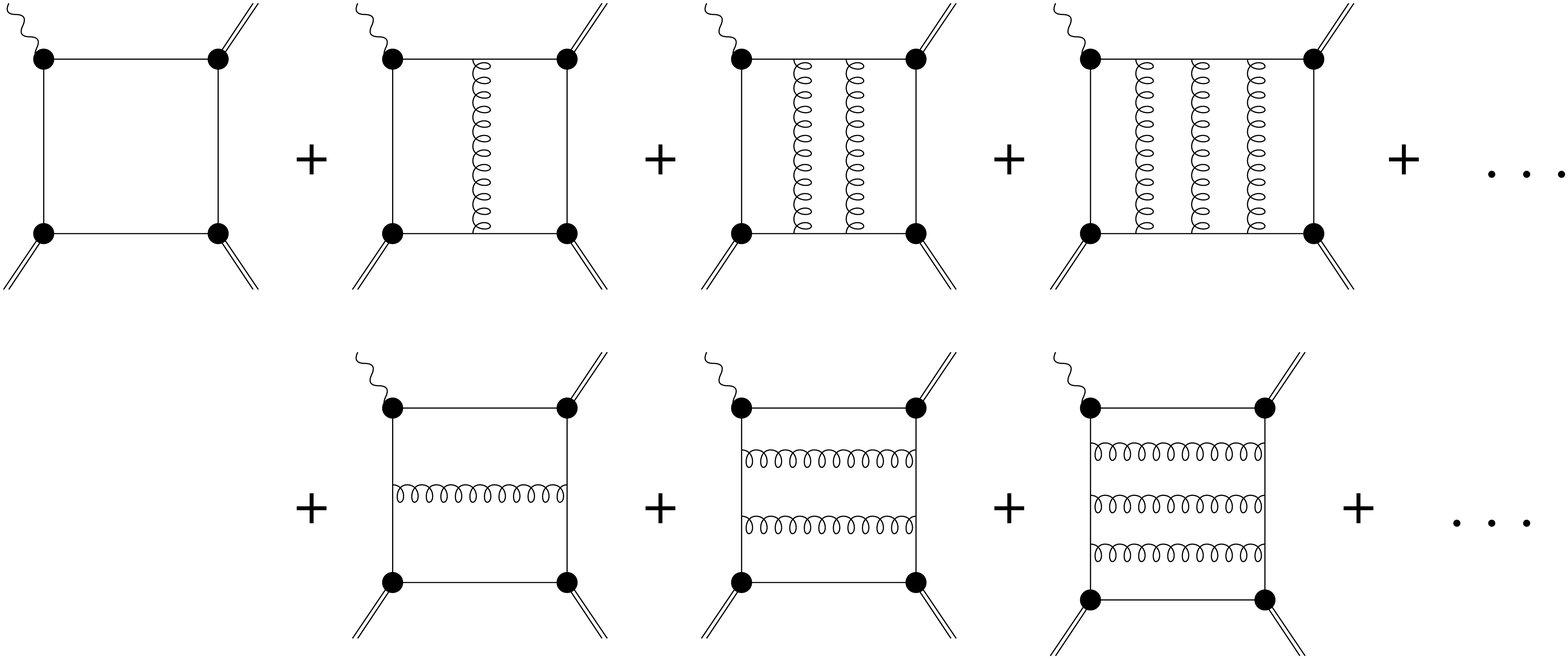}
\caption{\label{fig:beyond}
Gluon exchange diagrams needed to correctly describe $\gamma$-$3\pi$
processes in the rainbow-ladder truncation of the set of DSEs.}
\end{figure}

A consistent treatment requires multiple $s$, $t$ and $u$-channel
gluon exchange, as depicted in Fig.~\ref{fig:beyond} for one of the
six configurations.
This truncation scheme is equivalent to summing all planar diagrams
ignoring gluon self-interactions (triple and quartic gluon vertices).
The summation of the infinite set of ladder diagrams entails solving
an inhomogeneous BSE of the type
\begin{eqnarray}
\lefteqn{ G(p,P_i,P_j) =  G_0(p,P_i,P_j) +
	\int^\Lambda_q \!{\cal G}(k^2)\, D_{\mu\nu}^{\rm free}(k) }
\nonumber \\ && {}
        \times
	\textstyle{\frac{\lambda^{\alpha}}{2}} \gamma_\mu
	S(q_+) \, G(q,P_i,P_j)\, S(q_-)
        \textstyle{\frac{\lambda^{\alpha}}{2}} \gamma_\nu
\label{eq:inhomBSE} \, ,
\end{eqnarray}
where now $q_+ = q + P_i$, $q_- = q - P_j$ and $G_0(p,P_i,P_j)$ is
an inhomogeneous term of the form $\Gamma_\pi(p_+,p) \, S(p) \,
\Gamma_\pi(p,p_-)$.  For example, for the $s$-channel gluon exchange
diagrams in Fig.~\ref{fig:beyond} with $i=4$ and $j=3$
\begin{eqnarray}
G_0(p,P_4,P_3) =
	\Gamma_\pi(p+P_4,p) \, S(p) \, \Gamma_\pi(p,p-P_3) \,.
\end{eqnarray}
The corresponding amplitude for the sum of the GIA and the
$s$-channel gluon exchange diagrams is (recall, momentum
conservation dictates $P_2 = -Q - P_3 - P_4$)
\begin{eqnarray}
\lefteqn{a_\mu(P_2, P_3, P_4) =
	N_c \int_q^\Lambda\! {\rm Tr}\Big[
	S(k+P_4)G(k,P_4, P_3)}
\nonumber \\ && {} \times
 S(k-P_3) \Gamma^\gamma_\mu(k-P_3,k-Q-P_3)
\nonumber \\ && {} \times
	S(k-Q-P_3) \Gamma_\pi(k-Q-P_3,k+P_4)\Big] \, .
\end{eqnarray}
This treatment of the ladder diagrams is identical to the calculation
reported for $\pi$-$\pi$ scattering~\cite{Cotanch:2002vj}.

%%%%%%%%%%%%%%%%%%%%%%%%%%%%%%%%%%%%%%%%%%%%%%%%%%%%%%%%%%%%%%%%%%%%%%%%%%%%%
\section{\label{sec:symresults}
Numerical results for symmetric pion momenta}
In contrast to DSE calculations for three-body processes, a
significant computational effort is now necessary since $G$ is a
function of three independent variables: $p^2$, $p \cdot P_i$ and $p
\cdot P_j$, for fixed $P_i$ and $P_j$.  Discretizing these variables
on a three-dimensional grid, iteration is utilized starting with
$G_0$.  In the absence of singularities, convergence requires from 20
to 30 iterations.  Close to a pole from an intermediate meson bound
state, the number of required iterations increases dramatically.
Typically, per fixed set of external variables ($Q^2$, $s$, $t$ and
$u$) away from intermediate meson poles, of order 20 CPU hours are
needed to compute the form factor to an estimated precision of 2\%.
Our codes run on a parallel supercomputer (IBM SP) using 16 to 256
processors.  Taking symmetric pion momenta reduces the necessary
computer time by at least a factor of three.

\subsection{\label{sec:model}
Model parameters}
The model interaction and parameters developed in
Ref.~\cite{Maris:1999nt} for the masses and decay constants of the
light pseudoscalar and vector mesons are used for our numerical
analysis.  The effective $\bar q q$ interaction is
\begin{eqnarray}
\label{gvk2}
\frac{{\cal G}(k^2)}{k^2} &=&
        \frac{4\pi^2\, D \,k^2}{\omega^6} \, {\rm e}^{-k^2/\omega^2}
\nonumber \\ && {}
        + \frac{ 4\pi^2\, \gamma_m \; {\cal F}(k^2)}
        {\textstyle{\frac{1}{2}} \ln\left[\tau +
        \left(1 + k^2/\Lambda_{\rm QCD}^2\right)^2\right]} \;,
\end{eqnarray}
with \mbox{$\gamma_m=12/(33-2N_f)$, $N_f=4$}, \mbox{$\Lambda_{\rm QCD}
= 0.234\,{\rm GeV}$}, \mbox{$\tau={\rm e}^2-1$}, \mbox{${\cal F}(s)=(1
- \exp\frac{-s}{4 m_t^2})/s$} and \mbox{$m_t=0.5\,{\rm GeV}$}. The
renormalization scale, \mbox{$\mu=19\,{\rm GeV}$}, is within the
perturbative domain~\cite{Maris:1997tm,Maris:1999nt}.  The degenerate
$u/d$ quark mass, $m_{u=d}^{\mu=1 {\rm GeV}} = 5.5 \, {\rm MeV}$, and
remaining parameters, \mbox{$\omega = 0.4\,{\rm GeV}$},
\mbox{$D=0.93\,{\rm GeV}^2$}, were determined by fitting $m_\pi$,
$f_{\pi}$ and the chiral condensate.  Table~\ref{tab:model} summarizes
fitted, $\pi$, and predicted, $\rho$ and $\omega$, masses which are in
good agreement with observation~\cite{Maris:1999nt}.  The model's
predictions are also in very good agreement with the most recent JLab
data~\cite{Volmer:2000ek} for the $\pi^{\pm}$ form
factor~\cite{Maris:2000sk}, $F_\pi(Q^2)$, and various other pion,
$\rho$ and $\omega$
observables~\cite{Maris:2003vk,Maris:2002mz,Jarecke:2002xd}, some of
which are listed in Table~\ref{tab:model} as well.  In addition to a
light pseudoscalar and vector meson, this model also exhibits a light
scalar meson, putatively identified as the $\sigma$
meson~\cite{Cotanch:2002vj,Maris:2000ig}.
\begin{table}[h]%[H] add [H] placement to break table across pages
\caption{\label{tab:model}
Selection of calculated and measured meson masses and coupling constants.}
\begin{ruledtabular}
\begin{tabular}{l|llr}
        & experiment~\protect\cite{Groom:2000in}& calculated	&\\
        & (estimates)	& ($^\dagger$ fitted) 		&\\ \hline
$m_{u=d}^{\mu=1 {\rm GeV}}$ &	5 - 10 MeV	& 5.5 MeV     	&\\
- $\langle \bar q q \rangle^0_{\mu}$
                & (236 MeV)$^3$ & (241 MeV$^\dagger$)$^3$ &\\
$m_\pi$         &  138.5 MeV &   138$^\dagger$ MeV	&\\
$f_\pi$		&  92.4 MeV  &	 92.5$^\dagger$ MeV	&\\
$m_\rho$        &  771 MeV   &   742 MeV	&\\
$g_\rho$	&  5.01	     &    5.07		&\\
$g_{\rho\pi\pi}$&  6.02	     &    5.14		&\\
$m_\omega$      &  783 MeV   &   742 MeV	&\\
$g_\omega$	&  17.06     &	 15.2		&\\
$m_\sigma$      & 400$\sim$1200 MeV &  670 MeV	&\\
\end{tabular}
\end{ruledtabular}
\end{table}

\subsection{\label{sec:quarkmass}
Quark mass dependence of {$F^{3\pi}$} }
For symmetric pion momenta, $s=t=u$, Eq.~(\ref{eq:perm}) simplifies to
\begin{eqnarray}
 	F^{3\pi}(s,s,s) &=& e \; f(s,s,s) \,.
\end{eqnarray}
Since this reduces the necessary computer time significantly, we use
this case to demonstrate that our scheme is in agreement with the
low-energy theorem.

\begin{figure}[h]
\includegraphics[width=8cm]{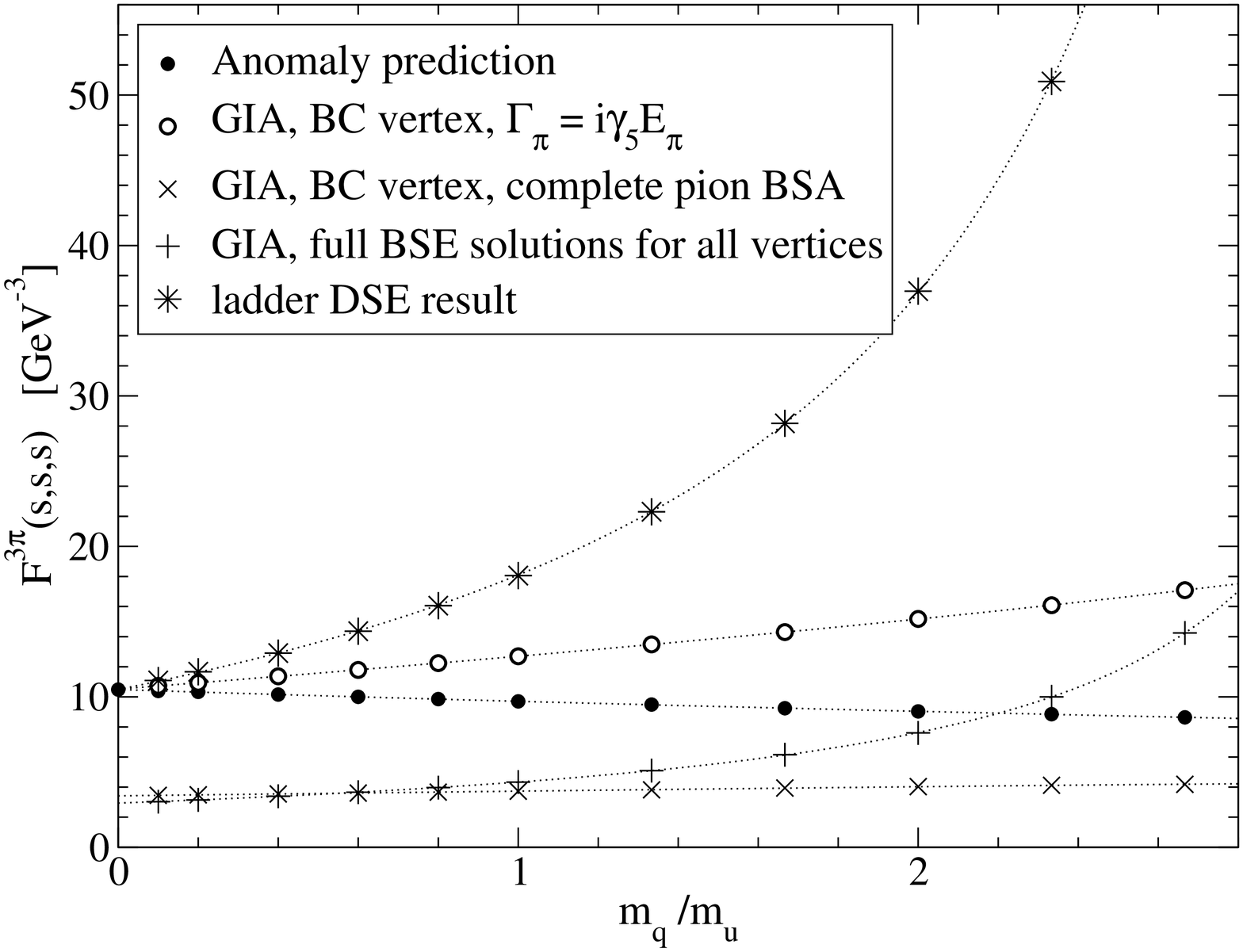}
\includegraphics[width=8cm]{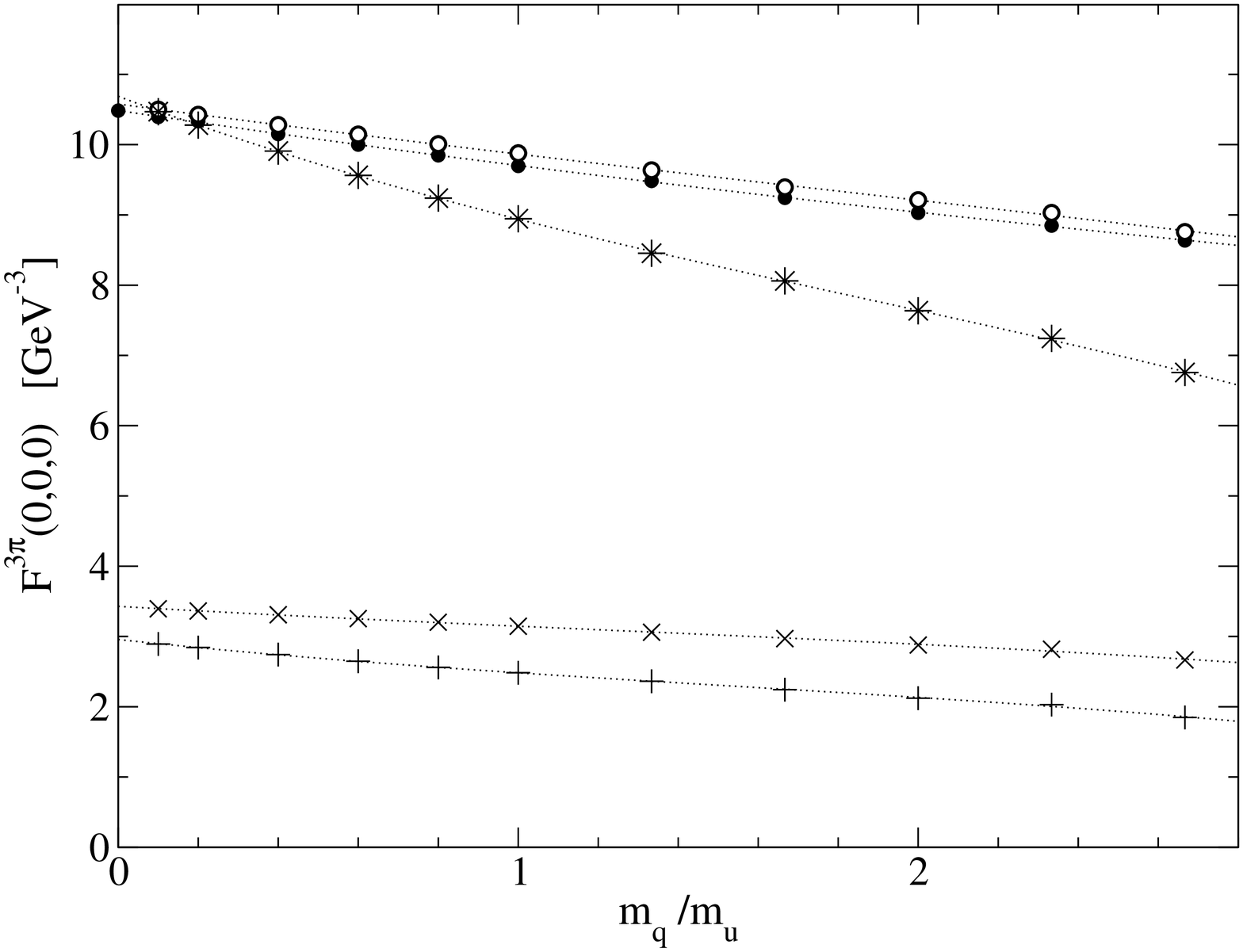}
\caption{\label{fig:chiral}
Numerical results for the symmetric form factor
$F^{3\pi}(4m_\pi^2,4m_\pi^2,4m_\pi^2)$ (top) and $F^{3\pi}(0,0,0)$
(bottom) as a function of the current quark mass $m_q$ divided by
the model up/down quark mass.  The curves are fits to our numerical
results (represented by the symbols) to guide the eye.}
\end{figure}

Note that the low-energy prediction, Eq.~(\ref{eq:chiralanomaly}), corresponds
to  threshold  production for three massless pions by an on-shell photon.
However, for physical pions, the threshold for three-pion production
is $s=t=u=4\,m_\pi^2$.  On the other hand, for equal pion momenta,
Eq.~(\ref{eq:MandelQ2}) reduces to $Q^2 = 3 (m_\pi^2 - s)$.  Thus it
is impossible to have on-shell photons ($Q^2=0$) producing three
physical pions.  Here we consider two limits, namely, threshold for
three-pion production by timelike photons,
$F^{3\pi}(4m_\pi^2,4m_\pi^2,4m_\pi^2)$ with $Q^2 = -9\,m_\pi^2$, and
$F^{3\pi}(0,0,0)$ with spacelike photon momenta $Q^2 = 3\,m_\pi^2$.
In the chiral limit, both should approach ${\hat F}^{3\pi}(0,0,0)$.
Our ladder DSE result for $F^{3\pi}$ is presented in Fig.~\ref{fig:chiral},
together with  results obtained using the GIA in combination with different
approximations for the vertices.  There is a small dependence of the
anomaly prediction on the current quark mass due to the fact that
$f_\pi$ depends on this mass.  To avoid confusion with
Eq.~(\ref{eq:chiralanomaly}), we denote the (mass-dependent) anomaly
prediction by
\begin{eqnarray}
  F_{\hbox{\scriptsize anomaly}}^{3\pi}(0,0,0) &=&
		\frac{e}{4 \pi^2 f_\pi^3}  \ .
\label{eq:pionanomaly}
\end{eqnarray}
For the physical current quark mass $f_\pi=92.4\;{\rm MeV}$ and
$F_{\hbox{\scriptsize anomaly}}^{3\pi}(0,0,0)=9.72\;{\rm GeV}^{-3}$.

Our ladder DSE approach does indeed reproduce the correct result in
the chiral limit.  The strong dependence of the form factor at
threshold, $F^{3\pi}(4m_\pi^2,4m_\pi^2,4m_\pi^2)$ with $Q^2 =
-9\,m_\pi^2$, is related to the $\omega$ pole in the quark-photon
vertex at timelike photon momentum $Q^2 = -m_\omega^2$.  The form
factor $F^{3\pi}(0,0,0)$ with spacelike $Q^2 = 3\,m_\pi^2$ is far less
dependent on the quark mass.  For on-shell photons, $Q^2 = 0$,
$F^{3\pi}(m_\pi^2,m_\pi^2,m_\pi^2)$ is essentially independent of the
current quark (and pion) mass.

The combination of GIA, BC ansatz for the quark-photon vertex and a
purely pseudoscalar BSA also reproduces the correct chiral limit.
However, the functional behavior approaching the chiral limit value is
quite different from our scheme.  This difference is predominantly due
to the lack of intermediate meson states, in particular the $\omega$,
which is not incorporated in the BC ansatz.  We address the importance
of intermediate meson states in Sec.~\ref{sec:mesonexchange}.  In
Fig.~\ref{fig:chiral} also note that the GIA with the BC ansatz and
full pion BSA including the pseudovector components $F$ and $G$ 
(see Eq.~(\ref{eq:piBSA})), does not produce the correct chiral
limit. Similarly, the GIA, with complete ladder BSE solutions for 
all vertices but without the additional ladder diagrams (see
Fig.~\ref{fig:beyond}), does not yield the anomaly prediction.

\subsection{\label{sec:photmom}
Dependence on the photon momentum }
\begin{figure}[h]
\includegraphics[width=8.5cm]{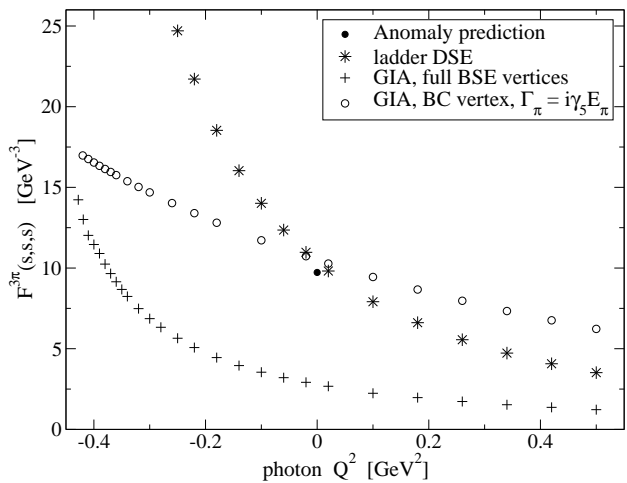}
\caption{\label{fig:Q2dep}
Numerical results for the symmetric form factor $F^{3\pi}(s,s,s)$ as
function of $Q^2$.}
\end{figure}

The dependence of the form factor on the photon momentum is shown in
Fig.~\ref{fig:Q2dep} for both timelike and spacelike $Q^2$, using the
model quark mass corresponding to physical pions.  Again, all three
on-shell pions have the same energy, $s=t=u=m_{\pi}^2 - Q^2/3$, and
the threshold for three-pion production is $Q^2 = -9 m_\pi^2\approx
-0.172~\hbox{GeV}^2$.

Our result is in good agreement with the chiral anomaly prediction, as
is the combination of the GIA and the BC ansatz for the quark-photon
vertex using a purely pseudoscalar BSA.  Now the difference between
the latter approach and our ladder DSE result is more apparent as our
result displays a much stronger $Q^2$-dependence.  In particular, our
result indicates a pole-like singularity for timelike $Q^2$, which can
be attributed to the presence of a vector meson pole in the dressed
quark-photon vertex.  This singularity is evident in both the GIA and
our full calculation, but not in the calculation using the BC ansatz.
Again, the GIA does not reproduce the anomaly prediction without the
additional ladder diagrams of Fig.~\ref{fig:beyond}.

%%%%%%%%%%%%%%%%%%%%%%%%%%%%%%%%%%%%%%%%%%%%%%%%%%%%%%%%%%%%%%%%%%%%%%%%%%%%%
\section{\label{sec:mesonexchange}
Comparison to meson exchange models}
The ladder DSE result clearly indicates a pole-like singularity in the
timelike $Q^2$ region, due to the presence of a vector meson pole in
the dressed quark-photon vertex.  In addition to this pole, there are
pole-like singularities from intermediate meson exchanges in the
$s$, $t$ and $u$ channels, similar to the scalar and vector meson
exchange contributions in $\pi$-$\pi$
scattering~\cite{Cotanch:2002vj}.  Here however, the scalar meson
exchange contribution is forbidden by spin/parity conservation.  Thus
only vector meson intermediate states contribute to $\gamma$-$3\pi$
processes.

\subsection{\label{subsec:VMD}
Vector meson dominance}
The effective vector meson dominance [VMD] Lagrangian is
\begin{eqnarray}
\cal{L}_{\text{int}} &=&
	g_{\rho\pi\pi} \vec\rho_{\mu} \cdot \vec\pi
		\times \nabla^\mu \vec\pi +
	g_{\omega \rho\pi} \epsilon^{\mu \nu \lambda \sigma}
		\nabla_{\mu}\omega_{\nu}\nabla_{\lambda}
		\vec\rho_{\sigma}\cdot \vec\pi
\nonumber \\ &&
	+ e(\frac{m^2_{\rho}}{g_{\rho}} \rho^0_{\mu} +
	\frac{m^2_{\omega}}{g_{\omega}} \omega_{\mu}) A^{\mu} 	 \ ,
\end{eqnarray}
where $\vec\pi, \vec\rho_{\sigma}, \omega_\nu$ and $A^\mu$ are the
$\pi$, $\rho$, $\omega$ and photon fields, $m_{\omega}$, $m_{\rho}$
are the $\omega$, $\rho$ masses and $g_{\omega\rho\pi}$,
$g_{\rho\pi\pi}$ are the $\omega \to \rho\pi$, $\rho \to \pi\pi$
coupling constants respectively.  With this Lagrangian the VMD form
factor for $\gamma \to \omega \to \rho\pi \to 3\pi$,
is~\cite{Rudaz:1974wt}
\begin{eqnarray}
\lefteqn{ F^{3\pi}_{VMD}(s,t,u) = \frac{2e m^2_{\omega}}{g_\omega}\;
 	\frac{g_{\omega\rho\pi} g_{\rho\pi\pi}}{Q^2 + m^2_\omega} }
	\nonumber \\ && \times
	\bigg( \frac{1}{m^2_{\rho} - s} +
		\frac{1}{m^2_{\rho} - t} + \frac{1}{m^2_{\rho} - u}
		\bigg) \, .
\label{eq:VMDff}
\end{eqnarray}
Note that this VMD form factor has vector meson poles in both the
photon momentum $Q^2$ and the Mandelstam variables $s$, $t$ and $u$.
The pole in $Q^2$ is associated with an intermediate $\omega$ meson,
whereas the poles in $s$, $t$ and $u$ are associated with $\rho$
intermediate states.

The measured $\omega, \rho \to e^+ e^-$ decay widths
\begin{eqnarray}
\Gamma^\omega_{ee} &=&
		\frac{4 \pi \alpha^2 m_{\omega}}{3\, g^2_{\omega}}
		= 0.60 \ \text{keV} \,,
\label{eq:omegadecay}  \\
\Gamma^{\rho}_{ee} &=&
		\frac{4 \pi \alpha^2 m_{\rho}}{3\, g^2_{\rho}}
		= 6.85 \ \text{keV} \,,
\label{eq:rhodecay}
\end{eqnarray}
determine the $\omega$-photon, $g_{\omega} = 17.06$, and
$\rho$-photon, $g_{\rho} = 5.01$, couplings which compare reasonably
with the model calculations of 15.2 and 5.07,
respectively~\cite{Maris:1999nt} (see also Table~\ref{tab:model}).
The hadronic coupling, $g_{\rho\pi\pi}$, was previously
predicted~\cite{Jarecke:2002xd} in this model to be 5.14 which agrees
favorably with the experimentally extracted value of 6.02.

Due to limited phase space, the $\omega \to \rho\pi$ decay is
forbidden, precluding direct determination of $g_{\omega\rho\pi}$.
However, this coupling can be related to the $\omega \to \pi^0 \gamma$
decay, again using VMD for $\omega \to \pi^0 \rho^0 \to \pi^0 \gamma$
\begin{eqnarray}
\Gamma^\omega_{\pi \gamma} &=& \frac{\alpha g^2_{\omega\rho\pi}
	(m^2_\omega - m^2_\pi)^3}{24 g^2_{\rho} m^3_\omega}  \, .
\end{eqnarray}
Using the measured width, $\Gamma^\omega_{\pi\gamma} = .734 \,{\rm
MeV}$, yields $g_{\omega\rho\pi} = 11.8 \, {\rm GeV}^{-1}$.  With
these parameters the VMD prediction for the $\pi^0 \to \gamma \gamma$
anomaly, via $\pi^0 \to \omega \rho \to \gamma \gamma$, is
\begin{eqnarray}
 F^{2\gamma}_{VMD}(0)
	&=& \frac{8 \pi \alpha g_{\omega\rho\pi}}{g_\rho g_\omega}
		= .0253 \, \text{GeV}^{-1} \, .
\end{eqnarray}
This is in excellent agreement with both the low-energy theorem,
$F^{2\gamma}(0) = \alpha / (\pi f_{\pi}) = .0251 \, {\rm GeV}^{-1}$,
and the experimental value, $.025 \pm 0.001 \, {\rm GeV}^{-1}$.

Returning to the $F^{3\pi}(s,t,u)$ form factor, the VMD prediction in
the chiral and zero momenta limits is
\begin{eqnarray}
 F^{3\pi}_{VMD}(0,0,0) &=& \frac{6\,e\,g_{\omega\rho\pi} g_{\rho\pi\pi}}
	{g_\omega m^2_{\rho}} = 12.6 \,\text{GeV}^{-3} \, .
\end{eqnarray}
This is above the low-energy theorem result,
Eq.~(\ref{eq:chiralanomaly}), but in potentially better agreement with
the observed value, $12.9 \pm 0.9 \pm 0.5 \, {\rm GeV}^{-3}$, measured
at higher energy (note that Ref.~\cite{Holstein:1996qj} quotes a
corrected experimental value of $11.9 \, {\rm GeV}^{-3}$).  By
reducing $g_{\rho\pi\pi}$ it is possible for the VMD form factor to
agree with the low-energy theorem, however this conflicts with the KSRF
relation~\cite{Kawarabayashi:1966kd,Riazuddin:1966sw}
\begin{eqnarray}
	g_{\rho\pi\pi} &=& m_{\rho}/(\sqrt{2} f_{\pi}) \,.
\label{eq:ksrf}
\end{eqnarray}
Further, the VMD result is not consistent with anomalous Ward
identities of Ref.~\cite{Aviv:1972hq}.  These two issues can be
resolved by adding a contact term to the Lagrangian
\begin{eqnarray}
 \cal{L}_{\text{contact}} &=& G_{\omega} \epsilon^{\mu \nu \lambda \sigma}
 \omega_{\mu}\nabla_{\nu} \vec\pi \cdot \nabla_{\lambda} \vec \pi \times
 \nabla_{\sigma} \vec \pi \ ,
\end{eqnarray}
with the coupling $G_\omega$ governing the direct $\omega \to 3\pi$
decay.  This gives the modified VMD
result~\cite{Rudaz:1984bz,Cohen:1989es}
\begin{eqnarray}
 \tilde{F}^{3\pi}_{VMD}(s,t,u) &=& \frac{-6\,e\, m^2_\omega}{g_\omega}
	\frac{G_\omega}{Q^2 + m^2_\omega} + F^{3\pi}_{VMD}(s,t,u)  \,.
\nonumber\\
\label{eq:modVMDff}
\end{eqnarray}

References~\cite{Rudaz:1984bz,Cohen:1989es} determine $G_\omega$ by
satisfying both Eq.~(\ref{eq:ksrf}) and the low-energy theorem for
$\gamma$-$3\pi$.  Further, these analyses also
invoke SU(2) flavor symmetry and universality using the relations
$g_\omega = 3 g_\rho = 3 g_{\rho\pi\pi}$, which are reasonable, to
obtain
\begin{eqnarray}
 G_\omega &=& \frac{g_{\rho\pi\pi}}{16 \, \pi^2 \; f^3_{\pi}} \,.
\end{eqnarray}
With these relations the modified form factor can be more succinctly
expressed in terms of only $f_\pi$, using Eq.~(\ref{eq:pionanomaly}),
instead of meson coupling constants
\begin{eqnarray}
\lefteqn{\tilde{F}^{3\pi}_{VMD}(s,t,u) = -\frac{1}{2} \frac{m^2_\omega \;
	F_{\hbox{\scriptsize anomaly}}^{3\pi}(0,0,0)}{Q^2 + m^2_{\omega}} }
\nonumber \\ &&
	\times\left[ 1 - \left(\frac {m^2_\rho}{m^2_\rho - s} +
	\frac{m^2_\rho}{m^2_\rho - t} + \frac{m^2_\rho}{m^2_\rho - u}
	\right) \right]  \ .
\label{eq:modVMD}
\end{eqnarray}

\subsection{\label{subsec:gam3pi}
Comparison of DSE and VMD for {$\gamma \to 3\pi$}}
Because of the established phenomenological success of VMD, it is
worth while to compare the two VMD form factors to our DSE result.  In
particular, it is of interest to ascertain that the DSE approach can
reproduce the resonance features of the VMD.  In doing so, consistency
mandates that the VMD phenomenological parameters be replaced by the
corresponding values calculated in the DSE model.  All necessary DSE
coupling constants, except for $g_{\omega\rho\pi}$ which is discussed
below, are listed in Table~\ref{tab:model}.
\begin{figure}[h]
\includegraphics[width=8.5cm]{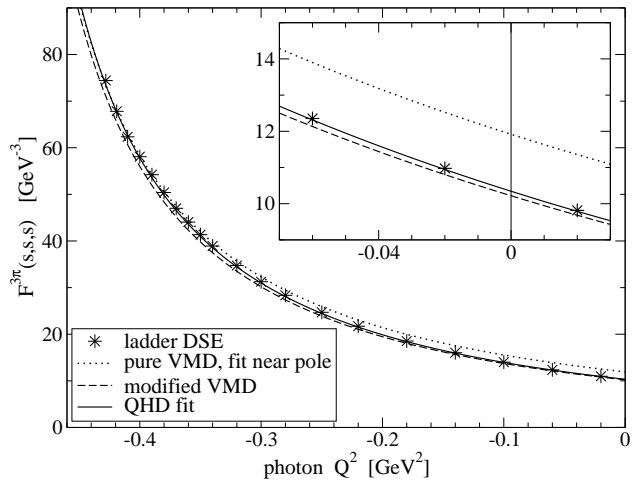}
\caption{\label{fig:vmd}
Comparison of the ladder DSE calculation (asterisk) to the DSE
parameterized pure (dotted) and modified (dashed) VMD form factors,
$F^{3\pi}(s,s,s)$, as function of $Q^2$.  The solid curve represents a
phenomenological QHD  fit to the DSE form factor.}
\end{figure}
In Fig. \ref{fig:vmd} the pure, Eq.~(\ref{eq:VMDff}), and modified,
Eq.~(\ref{eq:modVMD}), VMD form factors are compared to the DSE result
from $Q^2 \approx 0$ to the $\omega$ resonance region.

The ladder DSE and VMD results are in good qualitative agreement.  In
particular, the $\omega$ resonance exhibited in the VMD is well
reproduced by the ladder DSE approach.  Note that the DSE approach
naturally agrees with the low-energy theorem while only by
construction can the modified VMD form factor reproduce this result;
the pure VMD form factor does not reproduce the correct low-energy
limit.

Because of off-shell ambiguities, one can not consistently calculate
the unphysical $\omega \rho\pi$ amplitude.  However, by fitting the
pure VMD form factor, Eq.~(\ref{eq:VMDff}), to the DSE result near the
$\omega$ pole and using the calculated values of Table~\ref{tab:model}
for the remaining coupling constants in that equation, we extract the
value $g_{\omega\rho\pi} = 10.3 \, {\rm GeV}^{-1}$.  This is slightly
below the phenomenological VMD result of $11.8 \, {\rm GeV}^{-1}$ but
significantly lower than other theoretical values: SU(3) flavor symmetry,
$16 \, {\rm GeV}^{-1}$~\cite{RotelliScadron}; QCD sum rules, $15$ to
$17 \, {\rm GeV}^{-1}$~\cite{Eletsky:1983py}; light-cone sum rules,
$16 \, {\rm GeV}^{-1}$~\cite{Lublinsky:1997yf}.

In addition, Fig.~\ref{fig:vmd} also displays a form factor based on a
purely phenomenological quark-hadron model [QHD] incorporating both
contact interactions and meson-exchange contributions
\begin{eqnarray}
\lefteqn{ F^{3\pi}_{QHD}(s,t,u) \;=\;
	C_{\gamma3\pi} -
 \frac{6\,m^2_\omega\; C_{\omega3\pi} }{g_\omega\;(Q^2 + m^2_\omega)} }
\nonumber \\ && {}
 	- C_{\gamma\pi\rho}\;\left(
		\frac{m_\rho^2\;g_{\rho\pi\pi}}{m^2_\rho - s}
		+ \frac{m_\rho^2\;g_{\rho\pi\pi}}{m^2_\rho - t}
		+ \frac{m_\rho^2\;g_{\rho\pi\pi}}{m^2_\rho - u} \right)
\nonumber \\ && {}
	+ F^{3\pi}_{VMD}(s,t,u) \,.
\label{eq:QHDff}
\end{eqnarray}
The three coefficients $C$ are determined by fitting the ladder DSE
calculation at different $s$, $t$, $u$ and $Q^2$ values.  An accurate
fit over a large kinematical range is obtained with $C_{\gamma3\pi} =
0.248\;{\rm GeV}^{-3}$, $C_{\omega3\pi} = 5.94\;{\rm GeV}^{-3}$ and
$C_{\gamma\pi\rho} = 0.028\;{\rm GeV}^{-3}$.  This provides a useful
analytic representation of the DSE form factor.

\subsection{\label{subsec:gampitopipi}
Comparison of DSE and VMD for { $\gamma\pi\to\pi\pi$}}
Finally, we consider the process $\gamma\pi \to \pi\pi$.  It is
described by the same form factor $F^{3\pi}(s,t,u)$ as the process
$\gamma\to3\pi$, but the kinematical variables are different.  In
general, $s\neq t\neq u$, but for simplicity, we only consider the
more symmetric case $t=u$.  The kinematically allowed region is $s >
4\,m_\pi^2$ and $t,u < 0$.

\begin{figure}[h]
\includegraphics[width=8.5cm]{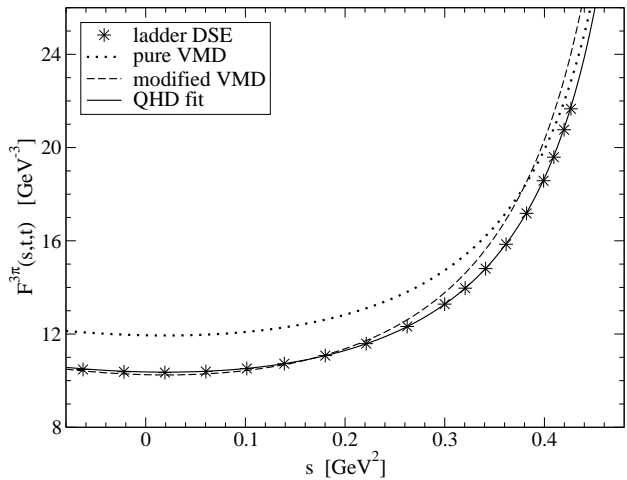}
\caption{\label{fig:sdep}
Comparison of the ladder DSE (asterisk), pure VMD (dotted), modified
VMD (dashed) and fitted QHD (solid) form factors, $F^{3\pi}(s,t,t)$,
as function of $s$ at $Q^2=0$.}
\end{figure}
In Fig.~\ref{fig:sdep} we compare our DSE results with the VMD and
phenomenological meson exchange form factors.  For fixed $Q^2=0$, the
$\omega$ meson does not appear as a pole in the form factor.  The
$\rho$ pole in the $s$ channel on the other hand is clearly
identifiable.  In addition, there are also poles in the unphysical
region, corresponding to intermediate $\rho$ mesons in the $t$- and
$u$-channels.

With the same parameters as in the previous subsection, the pure VMD
form factor agrees very well with the ladder DSE calculation near the
$\rho$ pole, but overestimates the form factor by about 15\% near
threshold.  The modified VMD form factor on the other hand agrees very
well with the DSE result near threshold, but starts to deviate for
increasing values of $s$.  The purely phenomenological QHD form factor
can describe the DSE form factor accurately both near threshold and in
the resonance region, not only at $Q^2 = 0$ but also at other values
of $Q^2$.  However, this is not surprising, since it was fitted to our
DSE result to provide an analytic representation of our results.

%%%%%%%%%%%%%%%%%%%%%%%%%%%%%%%%%%%%%%%%%%%%%%%%%%%%%%%%%%%%%%%%%%%%%%%%%%%%%
\section{\label{sec:expdata}
Comparison with experimental data}
Finally, for the benefit of experimentalist the same three form
factors are compared to the limited data.  We use the phenomenological
parameters in the VMD form factors permitting their most favorable
prediction.  In particular, the $\rho$ mass in Eqs.~(\ref{eq:VMDff})
and (\ref{eq:modVMD}) is replaced by $m_\rho - \frac{i}{2}\Gamma_\rho$
where $\Gamma_\rho = 149\;{\rm MeV}$ is the experimental width of the
$\rho$ meson.  This changes the $\rho$-meson pole singularity in
Fig.~\ref{fig:sdep} to a resonance peak with a finite maximum.  We
ignore the much smaller $\omega$ width since we consider the process
$\gamma\pi \to \pi\pi$ for on-shell photons ($Q^2 = 0$) only.

The ladder truncation of the inhomogeneous BSE
Eq.~(\ref{eq:inhomBSE}), on the other hand, has a real pole
singularity at the $\rho$-mass.  One would need to include two-pion
intermediate stated in the interaction kernel to consistently produce
a $\rho$-meson width.

\begin{figure}[h]
\includegraphics[width=8.5cm]{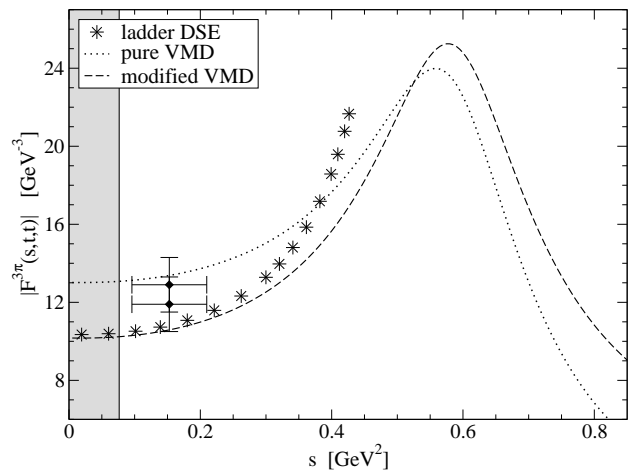}
\caption{\label{fig:expdata}
Comparison of DSE (asterisk), pure VMD (dotted), and modified VMD
(dashed) form factors, $F^{3\pi}(s,t,t)$, with data at $Q^2 = 0$.
The experimental coupling constants, masses and width are used in
both VMD form factors (the shaded region is unphysical).  The data
point is from Ref.~\protect\cite{Antipov:1987tp} (re-analysed in
Ref.~\protect\cite{Holstein:1996qj}); new experimental results from
JLab are anticipated~\protect\cite{Miskimen:private} in the region
$0.27\;{\rm GeV}^2 < s < 0.72\;{\rm GeV}^2$.}
\end{figure}
In Fig.~\ref{fig:expdata} we display the absolute value of both the
pure and the modified VMD form factors as functions of $s$, the total
energy of the two outgoing pions, for $t=u=(3m_\pi^2-s)/2$ and
on-shell photons.  For comparison, we also show the ladder DSE
calculation below the $\rho$-pole singularity.  The modified VMD form
factor appears to be the most realistic and agrees quite well with the
microscopic DSE calculation near threshold, $s = 4m_\pi^2$.  At larger
values of $s$ it reflects the $\rho$-meson width and therefore
exhibits a resonance peak rather than the pole singularity of the
ladder DSE calculation.  The large error bar on the experimental data
point precludes definitive conclusions but new, more precise
measurements are in progress which should provide useful insight and
model constraints.

%%%%%%%%%%%%%%%%%%%%%%%%%%%%%%%%%%%%%%%%%%%%%%%%%%%%%%%%%%%%%%%%%%%%%%%%%%%%%
\section{\label{sec:concl}
Conclusion}
In summary, the anomalous form factor for $\gamma$-$3\pi$ processes
has been calculated in a self-consistent Dyson--Schwinger,
Bethe--Salpeter formulation in the rainbow-ladder truncation.  The
significant finding is that to reproduce the fundamental low-energy
theorem, it is necessary to go beyond the generalized impulse
approximation.  Specifically, one has to include complete sets of
ladder diagrams in the $s$, $t$ and $u$ channels.

The ladder DSE result is in good agreement with vector meson
dominance.  The ladder dressing of the quark-photon vertex is
essential for generating the $\omega$ resonance in the
process $\gamma \to \omega \to 3\pi$.  Inclusion of the ladder
diagrams beyond the GIA (see Fig.~\ref{fig:beyond}) is necessary to
produce an intermediate $\rho$-meson state in the two-pion channel.
A meson-exchange form factor incorporating both contact interactions
and meson-exchange contributions, fitted to our DSE calculation,
provides a useful analytic representation of our results over a large
kinematical range.

Our predictions are also in agreement with the limited data near
threshold and await confrontation with additional and more precise
measurements currently in progress.  For larger values of $s$ we
expect the modified VMD form factor to be in better agreement with
experiment since it includes the $\rho$-meson width.  These results
should also be of interest to experimentalists investigating double
pion photo-production.

Future work will address calculating the width for $\omega \to 3\pi$
as well as the strangeness processes $\gamma \to K \bar{K} \pi$ and
$\gamma K \to \pi K$.  Longer term, this framework will be extended to
describe meson electro-production from a nucleon target.

%%%%%%%%%%%%%%%%%%%%%%%%%%%%%%%%%%%%%%%%%%%%%%%%%%%%%%%%%%%%%%%%%%%%%%%%%%%%%
% If you have acknowledgments, this puts in the proper section head.
\begin{acknowledgments}
This work was supported by the Department of Energy under grant
DE-FG02-97ER41048.  Calculations were performed with resources
provided by the National Energy Research Scientific Computing Center
and the North Carolina Supercomputer Center.
\end{acknowledgments}

% Specify following sections are appendices. Use \appendix* if there
% only one appendix.
%\appendix
%\appendix*
%
%%%%%%%%%%%%%%%%%%%%%%%%%%%%%%%%%%%%%%%%%%%%%%%%%%%%%%%%%%%%%%%%%%%%%%%%%%%%%
% Create the reference section using BibTeX:
\bibliography{refsPM,refsPCT,refsCDR,refs,refsgam3pi}
%\bibliography{refsgampi}

%%%%%%%%%%%%%%%%%%%%%%%%%%%%%%%%%%%%%%%%%%%%%%%%%%%%%%%%%%%%%%%%%%%%%%%%%%%%%
\end{document}